\documentclass{aa}
\usepackage{graphicx}
\usepackage{natbib}
\usepackage{txfonts}
\bibpunct{(}{)}{;}{a}{}{,}

\newcommand\changed{}
\newcommand\changedd{}

\usepackage{fix2col} 

\begin{document}

\title{Ellipticity and Deviations from Orthogonality in the Polarization Modes of PSR B0329+54}
\titlerunning{Ellipticity and Deviations from OPM in PSR B0329+54}
\author{R.~T. Edwards\inst{1} 
   \and B.~W. Stappers\inst{2,1}}
\date{}
\institute{Astronomical Institute ``Anton Pannekoek'', 
        University of Amsterdam,
        Kruislaan 403, 1098 SJ Amsterdam, The Netherlands 
  \and
   Stichting ASTRON, Postbus 2, 7990 AA Dwingeloo, The Netherlands}
\offprints{R.~T. Edwards,\\ \email{redwards@science.uva.nl}}

\abstract{We report on an analysis of the polarization of single
pulses of PSR B0329+54 at 328~MHz. We find that the distribution of
polarization orientations in the central component diverges strongly
from the standard picture of orthogonal polarization modes (OPMs),
making a remarkable partial annulus on the Poincar\'{e} sphere.  A
second, tightly clustered region of density appears in the opposite
hemisphere, at a point antipodal to the centre of the annulus.  We
argue that this can be understood in terms of birefringent alterations
in the relative phase of two elliptically polarized propagation modes
in the pulsar magnetosphere (i.e. generalised Faraday rotation). The
ellipticity of the modes implies a significant charge density in the
plasma, while the presence of both senses of circular polarization,
and the fact that only one mode shows the effect, supports the view
that refracted ordinary-mode rays are involved in the production of
the annulus.  At other pulse longitudes the polarization (including
the circular component) is broadly consistent with an origin in
elliptical OPMs, shown here quantitatively for the first time, however
considerable non-orthogonal contributions serve to broaden the
orientation distribution in an isotropic manner.  
}
\date{Received 10 February 2004 / Accepted 26 March 2004}
\maketitle
\section{Introduction}
\label{sec:intro}
Whenever sufficient sensitivity is available, radio pulsar emissions
are seen to be rich in phenomenology. Their polarization is no
exception to this rule. The dependence of linear polarization position
angle on pulse longitude (i.e. rotational phase) can, for some
pulsars, be explained as arising in the vicinity of the magnetic 
pole, polarized linearly at the position angle of the sky projection
of magnetic field lines \citep{rc69a}. For other pulsars this is not
the case, and for some of these the distribution of position angles
(PAs) in individual pulses has been shown to be bimodal about two
values separated by 90$\degr$ --- so-called orthogonal polarization
modes (OPMs; e.g. \citealt{mth75,brc76,crb78, br80,scr+84,scw+84}).
Many pulsars also show $90\degr$ ``jumps'' in their position angle
profiles, a fact that received explanation in the discovery of OPMs
through a longitude dependence of the relative intensities of the
modes, which themselves tend to have position angle swings
consistent with the magnetic pole model of \citet{rc69a}. These
studies of OPMs also found evidence for deviations from orthogonality
in the fact that PA distributions were broader than expected and/or
their peaks were not separated by $90\degr$.

Attempts have been made to explain these deviations from orthogonality
by means of superposition of two modes that are not orthogonal due to
origins on different field lines and subsequent birefringent
refraction (e.g. \citealt{scr+84,mck03a}), by means of the
superposition of a range of modal orientations arising from a
distribution of field lines that are visible due to their finite beam
width \citep{gl95}, or {\changed due to the presence of two instantaneously
orthogonal modes, the orientation of which varies with time due to
coherent wave coupling effects}
(e.g. \citealt{cr79,lp99}). Of these, only the first and third
scenarios have been quantitatively tested. \citet{mck03a} showed that
the distribution of PAs, and the shape of the average PA curve of PSR
B2016+28 are consistent with the superposition of two non-orthogonal
modes.  On the other hand, \citet{pet03} showed that the PA and
circular polarization of the average pulse profiles of PSR B0355+54
and PSR B0628$-$28 are consistent with the predictions of \citet{lp99}
for alterations in the polarization due to coherent effects. However,
neither work examines whether the data are also consistent with other
models, and indeed the statistics considered in each case are not well
suited to distinguishing between models.

The question of the origin of circular polarization in pulsars is also
not addressed by the magnetic pole model, in which the polarization is
expected to be linear due to the high magnetic field strength.  The
fact that orthogonal linear polarization states have been found to be
associated with opposite signs of circular polarization leads
naturally to the suggestion that the modes are in fact elliptically
polarized orthogonal modes \citep{crb78}, although to our knowledge,
to date no observational tests of the expected proportionality between
linear and circular polarization under this hypothesis have been
made. In explaining the origin of elliptical OPMs, most authors point
to the so-called polarization-limiting region (PLR), where
birefringent propagation effects no longer significantly alter the
relative phase of the modes, as the determinant of the observed
polarization \citep{cr79}.  It has been proposed that either the
propagation modes themselves are elliptical at this point,
requiring a net charge density in the plasma \citep{cr79,am82, hlk98},
or that weak birefringence in the vicinity of the PLR causes initially
linearly polarized rays to suffer changes in their polarization if
their position angle deviates from that of the local linear modes, for
example due to rotational aberration or refraction
\citep{cr79,lp99,pl00,pet01,pet03}.

In this work we make a detailed study of the polarization of single
pulses from PSR B0329+54, employing new techniques in the hope
of placing greater constraints on the nature and origin of ellipticity
and non-orthogonality of pulsar polarization. 

\section{Methods of Analysis}
\subsection{Stokes Parameters of OPM}
\label{sec:stokes}
{\changed Since the focus of this study is the inconsistency of
observations with a simple model of superposed OPMs, it is pertinent
to begin with a clear picture of the features expected under such a
model, before discussing ways in which the observations may deviate
from it.}  The distribution of Stokes parameters expected under the
incoherent superposition of two elliptical OPMs, and techniques for
reconstruction of the modal intensities have been considered in detail
by \citet{ms00}.We briefly re-iterate in a more compact vector form
before considering the fluctuation statistics.

The modes are taken to be completely polarized, with fixed position
angles and degrees of circular polarization at a given pulse
longitude. This means that the vector $\vec{p}_i=\left(Q\;U\;V\right)^T$
for a given mode $i$ always has the same orientation, regardless of its
length (i.e. the intensity of the mode).  The condition of
orthogonality of the electric field vectors requires that the
polarization states be antiparallel in the Poincar\'{e} sphere
($\vec{p}$-space). Since the Stokes 4-vector of the incoherent sum of
two polarization states is simply the sum of their respective Stokes
vectors, all resultant states must lie on the line defined by the
modal orientation.  The observed Stokes parameters at any instant
can be written as:
\begin{eqnarray}
I &=& I_1 + I_2  \label{eq:i1pi2}\\
\vec{p} &=& p_1 + p_2 \nonumber \\
        &=& (I_1 - I_2)\vec{1_m} \label{eq:i1mi2},
\end{eqnarray}
\noindent where the contributions from the two modes are denoted with
numerical subscripts, with $\vec{1_m}$ being a unit vector in the
direction associated with the mode denoted ``1''.  

The procedure of
``mode separation'' (determination of modal intensities) follows
directly as:
\begin{eqnarray}
I_1 &=& (I+|\vec{p}|)/2 \label{eq:i1} \\
I_2 &=& (I-|\vec{p}|)/2, \label{eq:i2}
\end{eqnarray}
\noindent while the modal orientation is simply given by $\vec{1}_m =
\vec{p}/|\vec{p}|$. Here we have explicitly identified mode 1 as the
greater in intensity. Note that $|\vec{p}|$ is subject to bias in the
presence of measurement noise. Although this can be corrected, if some
alternative means is available for determining the modal orientation,
since components of $\vec{p}$ orthogonal to $\vec{1}_m$ contribute
only noise to $|\vec{p}|$ one may eliminate the bias and improve
sensitivity by substituting $\vec{p}\cdot \vec{1}_m$ for $|\vec{p}|$.
Prior information on $\vec{1}_m$ could, for example, be obtained from
studies of the statistics of $\vec{p}$ (Sect.\
\ref{sec:pca}), or for mode-separation of individual pulses,
from the average profile.

By measuring appropriate statistics of the Stokes vectors, information
is available on the fluctuation statistics of the OPMs. Characterising
the latter by the modal variances $\sigma^2_1$ and $\sigma^2_2$ and
their covariance $\sigma_{12}$, one observes that
\begin{eqnarray}
\sigma^2_I &=& \sigma^2_1 + \sigma^2_2 + 2\sigma_{12} \\
\sigma^2_m &=& \sigma^2_1 + \sigma^2_2 - 2\sigma_{12}, 
\label{eq:sigmasqm}
\end{eqnarray}
\noindent where $\sigma^2_m$ is the variance of $\vec{p}\cdot
\vec{1}_m$. From this the covariance of the modes can be calculated as
\begin{equation}
\sigma_{12} = (\sigma^2_I-\sigma^2_m)/4 .
\label{eq:covar_modes}
\end{equation}
\noindent {\changed In practice the measured variances will be biased by
the presence of measurement noise, which can be estimated using the 
off-pulse covariances (and the pulsar intensity, if it contributes
appreciably to the system temperature), and subtracted.}

\subsection{Characterising the Distribution on the Poincar\'{e} Sphere}
\label{sec:distribution}
The defining feature of elliptical OPMs is the predicted constant
orientation of $\vec{p}$, regardless of pulse-to-pulse fluctuations in
$I$ and $|\vec{p}|$. This fact is put to good use in the display of
histograms of the position angle of linear polarization,
$\psi=\frac{1}{2}\tan^{-1} U/Q$, which at a given longitude can take
one of two allowed values offset by $90\degr$. In the presence of
instrumental noise these broaden to give the characteristic bimodal
distribution of OPM. The position angle distribution gives information
on the orthogonality of the linear component of the polarization, but
the question must also be asked whether the circular component is as
expected under elliptical OPM.  Attempts to answer this question
have been hampered in the past due to the use of inappropriate
statistics.  \citet{mck02} notes that the picture of elliptical OPM is
consistent with observations showing distributions of Stokes $V/I$
that are broad and centred near zero \citep{br80,scr+84} (or
equivalently, we note, that $\left<|V|\right>$ tends to greatly exceed
$|\left<V\right>|$; \citealt{kjm+03}). However, such a distribution
could not be considered a particularly distinctive feature of OPM and
could easily be produced even by a mechanism for the production of
circular polarization that is independent of the linear OPM
phenomenon.  A much more stringent test is to check that the observed
$V$ is, along with $Q$ and $U$, consistent with a constant orientation
of $\vec{p}$.  The natural complement to the position angle in this
regard is the ellipticity angle, $\chi =\frac{1}{2}\tan^{-1}
V/\sqrt{Q^2+U^2}$ : together they completely specify the orientation
of $\vec{p}$ via the spherical coordinate angles $2\psi, 2\chi$. The
distributions in both parameters should be bimodal under OPM (unless
one mode always dominates), unlike the distribution of $V/I$, 
{\changed which may be unimodal. }

While a measurement of the joint probability density function,
$f(\psi,\chi)$, contains sufficient information to detect the presence
of elliptical OPM, this choice of parameterisation is not ideal. This
is because a given solid angle element on the Poincar\'{e} sphere
subtends an area in $\psi,\chi$-space that itself depends on $\chi$.
Specifically,
\begin{equation}
{\rm d}\vec{a} = 4{\rm d}\psi{\rm d}\chi \cos 2\chi.
\end{equation}
\noindent This means that points closer to the equator ($\chi=0$) 
are given greater weight in the distribution than justified by their
density on the sphere (see also \citealt{mck03b}). This problem could
be circumvented by using $f(\psi,\chi)/\cos 2\chi$ instead, but the
problem remains that the spatial scale with which features in the 
distribution are represented in Cartesian $\psi,\chi$ coordinates
(for example in a plot) varies with $\chi$. 

A solution to this problem is to measure $f(2\psi, \sin 2\chi)$, which 
satisfies the equal-area condition
\begin{equation}
{\rm d}\vec{a} = {\rm d}(2\psi) {\rm d}(\sin 2\chi).
\end{equation}
\noindent This choice of parameterisation is known as Lambert's
cylindrical equal-area projection. In fact, any equal-area projection
has the desired properties of providing true solid-angle densities and
representative sizes for features. Lambert's cylindrical equal-area
projection suffers from severe shape distortions at both poles
($Q=U=0$, i.e. $\chi=\pm 45\degr$) and we shall only make use of it in
the case where one of the two dimensions is to be averaged in order to
use the other along with pulse longitude in a two-dimensional
display. That is, we supplement the traditional $\psi$ versus
longitude density display with a $\sin 2\chi$ versus longitude display
that gives information on the distribution of ellipticity as a
function of longitude.

To examine the distribution of polarization orientations at a single
pulse longitude, it is desirable to have minimum (zero) distortion at
both of the orientations associated with the OPMs. A cylindrical
equal-area projection with its equator containing the OPM orientations
would suffice, however the nature of deviations from OPM in PSR B0329+54
we describe below motivates the use of a projection where the distortion
is axisymmetric about the modal orientation. The only such projection
to also have the equal area property is Lambert's azimuthal equal
area projection, $f(x,y)$ where
\begin{eqnarray}
x &=& \rho\cos\lambda \\
y &=& \rho\sin\lambda \\
\rho &=& 2 \sin \left(\frac{\pi}{4} - \frac{\theta}{2}\right)
\end{eqnarray}
\noindent and $\lambda,\theta$ are the azimuth and latitude of
$\vec{p}$ in a rotated frame with $\vec{1}_m$ as its ``north'' pole.
This projection maps the entire sphere to a disc, with the ``north''
pole in the centre and the ``south'' pole distorted to a ring about
the circumference. In order to avoid this severe distortion, we
truncate the projection at $\theta < 0$ and display the southern
hemisphere in an adjacent projection with $-\vec{1}_m$ as its central
pole.

\subsection{Testing for Orthogonality}
\subsubsection{Eigenanalysis of the Covariance Matrix}
\label{sec:pca}
Previous studies of pulsar polarization have frequently revealed
position angle distributions with components that are significantly
broader than that expected due to instrumental noise alone
(e.g.\ \citealt{scr+84}), indicating
the presence of some non-orthogonal radiation.
 Deriving as it does
from the ratio of two noisy quantities ($Q$ and $U$), the linear
polarization position angle possesses complicated statistics,
depending not only on the statistics of measurement noise but also
on the statistics of variations in the linear polarization of the
source. Although much effort has been expended modelling them
(e.g. \citealt{mck03b}), it is clear that as a means of detecting
non-orthogonality, the PA distribution is less than ideal.  The
distribution of ellipticity angles suffers from the same problems, in
addition to those discussed in Sect.\ \ref{sec:distribution}. We suggest
instead a more robust test of the consistency of polarization
observations with an origin in OPM.

From the fact that OPMs contribute to the observed values of $\vec{p}$
only along a single vector, it follows that a simple and direct means
of detecting non-orthogonal emission is to check the statistics of the
components of $\vec{p}$ perpendicular to $\vec{1}_m$. In order to do
this, it is necessary to define a new orthogonal basis for $\vec{p}$,
which has $\vec{1}_m$ as one of the basis vectors. The question then
arises of what to use for a working value of $\vec{1}_m$, since when
the orthogonality of all states is in question, the mean
polarization vector may not be parallel to $\vec{1}_m$. Under the
assumption that the majority of the fluctuations in $\vec{p}$ are
directed along $\vec{1}_m$, a sensible choice for this vector is the
direction of greatest variance in $\vec{p}$, which is also the
least-squares estimate of the direction of fluctuations.  The method
of Principal Components Analysis (PCA; e.g. \citealt{jol86}) is
suggested as a means of finding this vector. PCA is based on the fact
that the set of eigenvectors of the covariance matrix of a
multivariate statistic represent an orthogonal basis in which the
variations in the different axes have zero covariance\footnote{We note
that a similar technique was independently described for OPM use by
\cite{mck04} while this paper was in the final stages of
preparation.}. That is, given the covariance matrix
\begin{equation}
\vec{C} = \left<\left(\vec{p}-\left<\vec{p}\right>\right)\left(\vec{p}-\left<\vec{p}\right>\right)^T\right> 
\label{eq:covar}
\end{equation}
\noindent where $\vec{p}$ is a column vector, 
there exists a set of 3 orthonormal vectors $\vec{e_i}$ 
(eigenvectors) and scalars $\lambda_i$ (eigenvalues) such that
\begin{equation}
\vec{C}\vec{e_i} = \lambda_i \vec{e_i}.
\end{equation}
\noindent It is easily shown that the covariance matrix is diagonal if the
vectors are expressed using the eigenvectors as a basis. That is:
\begin{eqnarray}
\vec{C'} &=& \left<\left[\vec{M}\left(\vec{p}-\left<\vec{p}\right>\right)\right]\left[\vec{M}\left(\vec{p}-\left<\vec{p}\right>\right)\right]^T\right> \\
        &=& \left(
\begin{array}{ccc}
\lambda_1 & 0 & 0 \\
0 & \lambda_2 & 0 \\
0 & 0 & \lambda_3
\end{array}
\right),
\end{eqnarray}
\noindent where $\vec{M}$ is a matrix with the eigenvectors as its rows.

Clearly, the eigenvalues are equal to the variances in the components
aligned with the corresponding vectors. The eigenvector with the
greatest eigenvalue corresponds to our choice of $\vec{1}_m$, while
the other vectors and their associated eigenvalues present a
convenient basis for detection and characterisation of non-orthogonal
radiation. In theory there may be fewer than three distinct
eigenvalues indicating axisymmetry (two eigenvalues) or isotropy (one
eigenvalue) in the directional variance, although the presence of
measurement noise makes this have zero probability in practice.
However, measurement noise also introduces bias to the covariance
matrix that should be corrected by subtraction of the covariance of
the noise, estimated from off-pulse longitudes and including also
the contribution of the pulsar to the system temperature if significant.
Where the intrinsic
covariance is of a similar magnitude to the uncertainty in the
off-pulse covariance, this may result in negative eigenvalues, which
presents a problem if $\lambda_i^{1/2}$ is to be used as a measure of
the scale of intensity fluctuations, and it must be accepted that
estimates will be unavailable in some longitude bins. This problem is
also familiar from, for example, bias-corrected estimates of the
linearly polarized intensity,
$L=(Q^2+U^2-\sigma^2_Q-\sigma^2_U)^{1/2}$.

Having measured the variance of fluctuations in components parallel to
and perpendicular to the modal orientation, it is useful to define a
measure of the degree of deviation from purely linear
fluctuations. After \citet{cp95}, we define the polarization entropy
as:
\begin{equation}
H = -\sum_{i=1}^3P_i\log_3 P_i \;,\; P_i = \frac{\lambda_i}{\sum_j \lambda_j},
\label{eq:entropy}
\end{equation}
\noindent where $\lambda_i$ is the $i$-th eigenvalue. This quantity
is confined to the range $[0,1]$, where $H=0$ corresponds to
fluctuations completely confined to one direction, and $H=1$ corresponds
to an isotropic distribution of $\vec{p}-\left<\vec{p}\right>$.

\subsubsection{Effects of Scintillation}
\label{sec:scint}
{\changed The methods of the previous section work from the covariance
of the observed signal, which in practice may reflect not only the
intrinsic variations of the pulsar, but also the effects of the
interstellar medium on the propagating signal. Of principle importance
is the variable effective gain of the interstellar medium induced by
two types of propagation effect, refractive and diffractive
scintillation.  For most pulsars, the time scale for variations due to
refractive scintillation is long ($\sim15$ days for PSR B0329+54;
\citealt{ssh+00}), so the effects can be neglected over short
observations.  Time scales for diffractive scintillation, on the other
hand, are much shorter. In some cases the time scale may be long
enough that one can simply limit the analysis to a segment of time
over which the flux density is constant (e.g. \citealt{mck04}), but in
other cases such intervals do not last long enough to sensitively
obtain representative statistics.  With a diffractive scintillation
time scale of $\sim148$~s at 328~MHz \citep{cor86}, this is the case
for the observations of PSR B0329+54 reported here.

In order to separate the (co-)variance induced by scintillation from
that intrinsic to the pulsar, we employ fluctuation spectral
techniques \citep{es03a}. We define the (longitude-resolved)
polarization spectral density tensor as the element-wise product of
the (discrete, vector) Fourier transform of
$\vec{p_j}-\left<\vec{p_j}\right>$ with its Hermitian transpose, where
$j$ indexes pulse number and the longitude dependence is implicit as
elsewhere in this work. That is,
\begin{eqnarray}
\vec{S_k} &=& \vec{P_k}\vec{P_k}^\dag, {\rm where} \label{eq:spec}\\ 
\vec{P_k} &=& \frac{1}{N}\sum_{j=1}^{N} e^{-2\pi ijk/N}\left(\vec{p_j}-\left<\vec{p_j}\right>\right),
\label{eq:vecspec}
\end{eqnarray}
\noindent $i=\sqrt{-1}$, $N$ is the number of pulses, and $^\dag$
denotes the Hermitian transpose. Here $k$ is a frequency index
(corresponding to a frequency of $k/N$ cycles per pulse period), which
is most easily interpreted if it is taken in the range $(-N/2,N/2]$.
For convenience, in practice the observation can be broken into
segments for which $\vec{S}$ is computed, and the results
averaged. The tensor $\vec{S}$ contains a wealth of information about
the polarization fluctuations, offering among other possibilities, a
means for distinguishing different sources of variation in $\vec{p}$
via their different fluctuation frequency structure, and a chance to
form the complex covariance matrix of the analytic signal and check
for periodic patterns of elliptical (rather than only linear) form
(Edwards 2004, in preparation).  In the present context we simply make
use of the fact that the sum of $\vec{S_k}$ over all $k$ gives, via
Parseval's theorem, the covariance matrix (Eq.\ \ref{eq:covar}), while
the covariance matrix of a bandpass-filtered version of the data can
be computed as a sum over a suitably limited range of $k$.

The effect of {\changedd scintillation} on the observed signal $\vec{p_j}$ is
a time-varying, multiplicative gain. The observed signal can be written
\begin{equation}
\vec{p_j} = \left(1+m\rho_j\right) \vec{p^i_j},
\label{eq:scintsig1}
\end{equation}
\noindent where $m$ is the modulation index due to scintillation,
$\rho_j$ is a realisation of a random process of zero mean and unit
variance, and $\vec{p^i}$ is the intrinsic signal. Breaking the
intrinsic signal into a mean contribution plus a variable
contribution, Eq. \ref{eq:scintsig1} can be expanded to give
\begin{equation}
\vec{p_j}-\left<\vec{p}\right> = \left(1+m\rho_j\right)\left(\vec{p^i_j} - \left<\vec{p^i}\right>\right)
  + m\rho_j\left<\vec{p^i}\right>.
\label{eq:scintsig2}
\end{equation}
Assuming that the instrinsic and scintillative variations are independent,
the covariance matrix (Eq.\ \ref{eq:covar}) becomes
\begin{equation}
\vec{C} = \left(1+m^2\right)\vec{C^i} + m^2\left<\vec{p^i}\right> \left<\vec{p^i}\right>^T .
\label{eq:covarscint}
\end{equation}
\noindent This shows that the uncorrected covariance matrix is biased
and scaled by the presence of scintillation\footnote{\changed As an
aside we note that the presence of scaling implies, surprisingly, that
scintillation can improve sensitivity to intrinsic variance, even when
the data contain an equal balance of scintillative amplification and
deamplification events.}. Replacing multiplications in the time domain
of Eq. \ref{eq:scintsig2} with convolutions in the frequency domain,
one finds that the vector spectrum of Eq. \ref{eq:vecspec} consists of
the sum of the spectrum of intrinsic variations from the mean, its
convolution with the spectrum of the scintillation variations, and the
convolution of the spectrum of the scintillation variations with the
spectrum of $\left<\vec{p^i}\right>$, i.e.\ a scaled sinc function at
zero frequency. The condition of independence implies that this sum
carries over to Eq. \ref{eq:spec} just as it did to Eq.\
\ref{eq:covarscint}.  The spectral response of the scintillation
variations has a characteristic width $1/\tau$ (where $\tau$ is the
scintillation timescale), corresponding to $\Delta k = NP/\tau$, where
$P$ is the pulse period. In the usual case that $P/\tau \ll 1$, we see
that the bias term in the observed covariance is confined to a small
fraction of the spectral density tensor around zero frequency. Under
the reasonable assumption that the fraction of intrinsic fluctuation
energy at these frequencies is small, this region can simply be
omitted in the sum over $k$ that is performed to form the covariance
matrix, to eliminate the bias. The remaining scale factor $1+m^2$ can
be obtained (given $\left<\vec{p}\right>$) from the complementary sum
of the spectral tensor over small $|k|$, which is over-constrained,
necessitating a fit.  In practice, since $m$ and (to first order)
$\rho_j$ are independent of pulse longitude, it can be obtained easily
and accurately by adding all on-pulse longitude bins in the Stokes I
signal and extracting the analogous quantity from the fluctuation
power spectrum of the result.
}

\section{Observations and Discussion}
\subsection{Observations}
We used the Westerbork Synthesis Radio Telescope with the PuMa pulsar
backend \citep{vkv02} to observe PSR B0329+54 in a 10~MHz band
centred at 328~MHz. PuMa was configured as a digital filterbank,
producing samples in all four Stokes parameters over 128 frequency
channels with a sample interval of 409.6 $\mu$s. We found that our
standard digitisation using 2 bits per sample per channel was
insufficient to avoid clipping the brightest pulses, and caution that
this may also have been a problem in previously published polarimetry
of this and other bright pulsars. This problem was avoided by
re-observation of the pulsar with 8 bits of digitisation. In offline
analysis the data were corrected for instrumental polarization effects
determined using the procedure described in the Appendix, followed by
removal of the frequency-dependent position angle rotation caused by
interstellar Faraday rotation, and samples were summed across all
frequency channels after correcting for delays due to interstellar
dispersion. The resultant time series was divided into segments
corresponding to the apparent pulse period to give an array of 16400
pulses in 1744 pulse longitude bins, 220 of which were used in
further analysis.

This pulsar is known to exhibit at least four main modes of emission,
each with a different pulse profile \citep{bmsh82}, two of which
are commonly seen at low frequencies. By forming pulse profiles
in sub-integrations of 100 pulses, we determined that the pulsar
underwent a mode change at pulse number $\sim 12800$. The profiles
formed by adding pulses 0--12500 and 13000--16400 are consistent
with the so-called ``normal'' and ``abnormal'' modes reported by
previous authors. 

\subsection{Distribution of Polarization Orientations}
\begin{figure*}
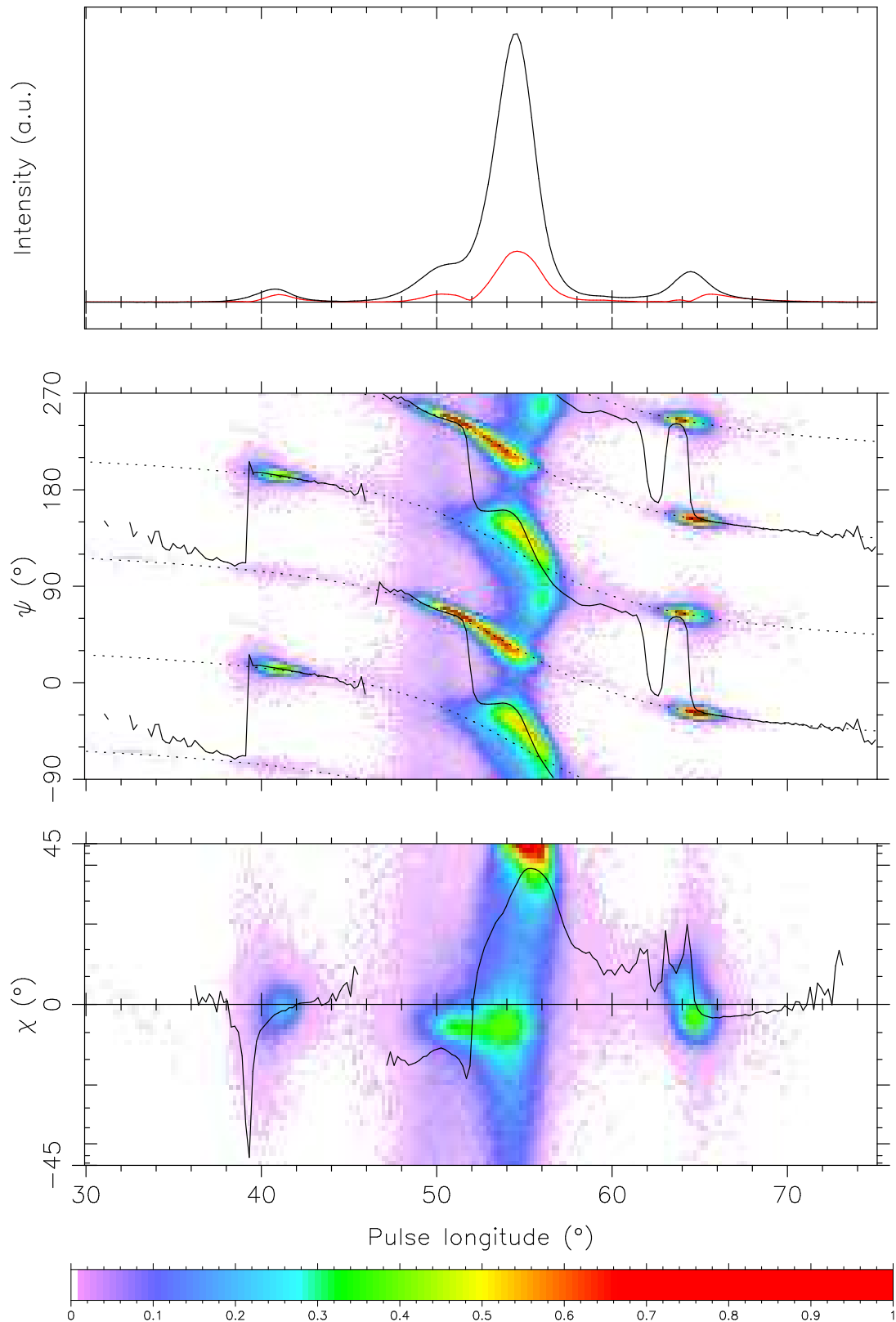

\centerline{\resizebox{0.9\hsize}{!}{\includegraphics{anghist.eps}}}
\vspace{-6mm}
\centerline{\resizebox{0.75\hsize}{!}{\includegraphics{gradient.eps}}}
\caption{Longitude-dependent polarization behaviour of PSR B0329+54 in
its normal profile mode at 328~MHz. Plotted are the mean total and
polarized intensity (black and red, top panel) and histograms of
position angle (middle panel, plotted twice for continuity) and $\sin
2\chi$ where $\chi$ is the ellipticity angle, along with the
corresponding parameters of the mean polarization vector (black
lines). {\changed For convenience the ordinate scale of the bottom
panel is labelled non-linearly in terms of $\chi$.}  The histograms
are normalised by the peak density and plotted in the colour scale
shown at the bottom.  The dashed line in the middle panel shows the
prediction under the magnetic pole model \citep{rc69a}, see text. The
smooth behaviour seen is intrinsic to the pulsar, the longitude bin
spacing is determined by the instrumental resolution. }
\label{fig:anghist}
\end{figure*}

As a first step in the characterisation of the distribution of
orientations of the polarization vector $\vec{p}$, we formed frequency
of occurrence histograms in position angle and $\sin 2\chi$
(Sect. \ref{sec:distribution}) as a function of pulse longitude.  To
reduce the effect of noise, only those samples with $|\vec{p}| > 10
\sigma_p$, where $\sigma_p$ is the quadrature mean of the variance
measured in Stokes $Q$, $U$ and $V$ in off-pulse longitude bins, were
included in the histograms. The result for the normal profile mode is
shown in Fig. \ref{fig:anghist}. 

\begin{figure*}
\begin{center}
\begin{tabular}{cc}
  \resizebox{0.43\hsize}{!}{\includegraphics{po1.eps}} &
  \resizebox{0.43\hsize}{!}{\includegraphics{po2.eps}} \\
\vspace{0.5mm} \\
  \resizebox{0.43\hsize}{!}{\includegraphics{po3.eps}}  &
  \resizebox{0.43\hsize}{!}{\includegraphics{po4.eps}} \\
\vspace{0.5mm} \\
  \resizebox{0.43\hsize}{!}{\includegraphics{po5.eps}}  &
  \resizebox{0.43\hsize}{!}{\includegraphics{po6.eps}} \\
\vspace{0.5mm} \\
  \resizebox{0.43\hsize}{!}{\includegraphics{po7.eps}}  &
  \resizebox{0.43\hsize}{!}{\includegraphics{po8.eps}} \\
 \multicolumn{2}{c}{\resizebox{0.9\hsize}{!}{\includegraphics{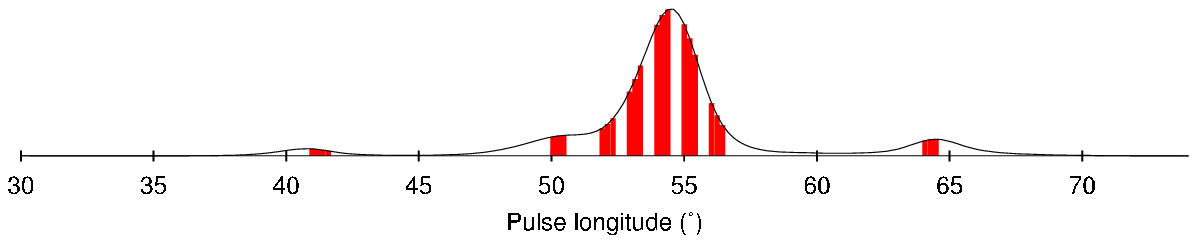}}}
\end{tabular}
\end{center}
\caption{Distribution of polarization orientations in eight longitude
intervals, marked in the pulse profile at the bottom and shown in 
left-to-right, top-to-bottom order. Each panel shows the distribution in
Lambert's azimuthal equal-area projection with poles set near to the
typical orientation of the mode showing the least scatter. The projection is
interrupted at the equator and plotted in two hemispheres, to give low
distortion near both poles, marked by asterisks. Lines of constant
$\psi$ and $\chi$ are marked for reference; meridians (constant $\psi$)
converge at $\chi=45\degr$ (pure
right-circular polarization)  near the top of the
projection. The colour scale is shown in Fig.\ \ref{fig:anghist},
each density being normalised by the peak in each longitude interval.}
\label{fig:po}
\end{figure*}

The histograms for the abnormal mode were qualitatively very similar
to those of the normal mode, upon which the remainder of the
discussion will focus. However, we note as an aside that the position
angle distributions appeared offset between the two profile
modes. Although the presence of non-OPM means that the magnetic pole
model of \citet{rc69a} is not strictly applicable, for the purpose of
quantifying the offset, after \citet{gl95} we attempted to fit the
model to position angles determined using local maxima in the
histograms near to a curve made by eye to approximately fit the less
distorted of the polarization modes (Fig. \ref{fig:anghist}).  We
found that for the abnormal mode, a fit to the fiducial position angle
parameter $\psi_0$ while using values determined from the normal mode
for other parameters performed much better than a fit where $\psi_0$
was fixed and all others allowed to vary (RMS deviation $3.13\degr$
versus $3.91\degr$), and was comparable to the best fit with all
parameters allowed to vary (RMS $3.06\degr$). We therefore conclude
that the offset is consistent with a shift in position angle of $\sim
-5.5\degr$, rather than a change in the apparent viewing geometry or a
longitude offset as might be induced by differential aberration and
retardation.

While the position angle distribution of Fig. \ref{fig:anghist} is
consistent with previously published results of lower resolution
showing quasi-orthogonal modes \citep{gl95}, the ellipticity
distribution shows features of a kind never seen before in any pulsar,
owing most likely to the fact that previous studies have used $V/I$
instead of ellipticity, causing OPM-related features to be washed out
due to fluctuations in $|\vec{p}|/I$. Most striking is the strong
right-circular polarization seen under the main central component,
which has no corresponding left-circular component of equal
ellipticity as would be expected if the circular polarization is due
to the OPM clearly seen in the position angle distribution
(Sects. \ref{sec:stokes} \& \ref{sec:distribution}).  Also of interest
is that the trailing component (longitude $\sim 64\degr$) has a
distribution that is roughly bimodal about the zero line, as expected
under elliptical OPM, while the leading component (longitude $\sim
41\degr$) appears to have a unimodal ellipticity distribution. Also
apparent is that polarized emission is occasionally detected in the
vicinity of pulse longitudes $33\degr$ and $70\degr$, corresponding to
the additional emission components detected in total intensity by
\citet{gg01}.

Much more intriguing behaviour is made apparent when the full
two-dimensional orientation distribution is considered for particular
longitude ranges. In Fig. \ref{fig:po} we display these distributions
averaged over several longitude intervals, using the
projection described in Sect.\ \ref{sec:distribution}. In what follows
we refer to the modes occurring in the left and right halves of each
projection as modes ``1'' and ``2'' respectively. Addressing the
distributions in longitude order, we see that the leading component is
consistent with purely linear OPM in mode 2, while by pulse longitude
$50\degr$ the modes have switched in dominance and become somewhat
elliptical. As the pulsar rotates, mode 2 begins to increase again
in strength, and apparently has a greater spread in its orientations
than mode 1. Over the course of the central component the distribution
associated with mode 2 deforms into an arc and eventually an almost
complete annulus, while mode 1 remains tightly distributed around
an elliptical orientation and eventually concedes dominance to mode 2.
Finally, in the trailing component the modes are of comparable
strength and distributed tightly around orthogonal elliptical
orientations. We discuss our interpretation of this remarkable
behaviour in Sect.\ \ref{sec:interp} but first discuss the remaining
observational results. 

\begin{figure}
\begin{center}
\resizebox{0.85\hsize}{!}{\includegraphics{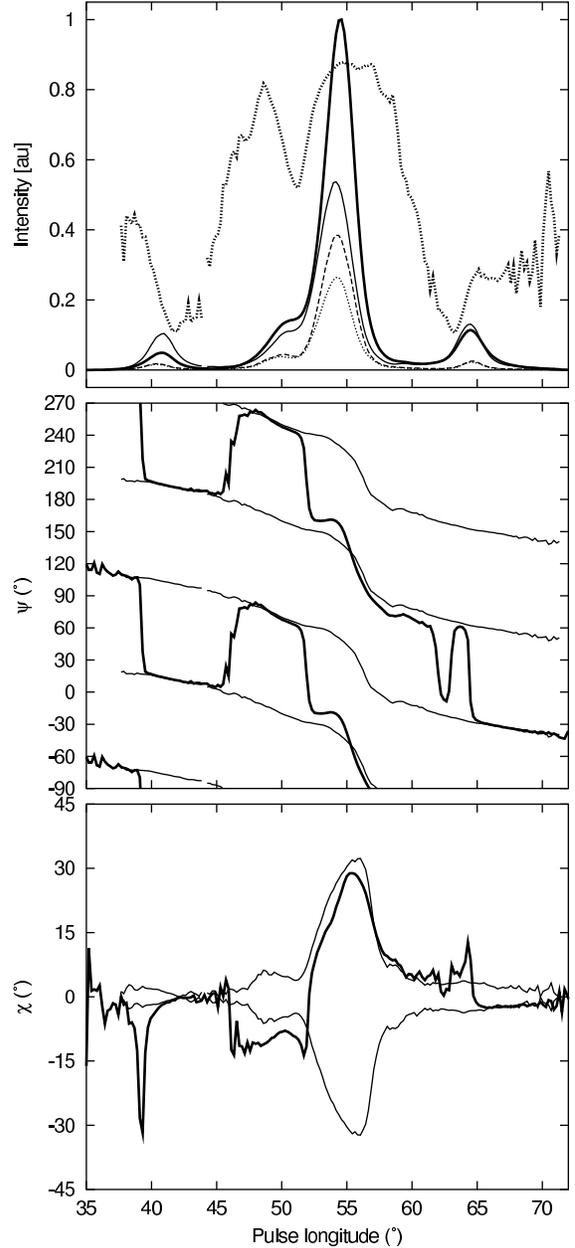}}
\end{center}
\caption{Results of eigenanalysis. The top panel shows the average
intensity profile (thick solid line), the square roots of the
eigenvalues (solid, dashed, dotted thin lines, in descending order of
value), and the polarization entropy (Eq.\ \ref{eq:entropy}; thick dotted
line). The middle panel
shows the position angle of the mean polarization vector (thick line,
plotted repeatedly at offsets of $180\degr$ for continuity) and of the
first eigenvector (thin line, plotted repeatedly at offsets of
$90\degr$ for continuity under the equivalence of antiparallel vectors
as eigenvectors).  The bottom panel shows the ellipticity angle of the
mean polarization vector (thick line) along with the ellipticity angle
of the first eigenvector, plotted twice with opposite signs.}
\label{fig:pca}
\end{figure}

\subsection{Covariance Analysis}
The results of the previous section leave no doubt that the central
component shows strong deviation from the behaviour expected under
OPM. The case of the leading and trailing components is more difficult
to assess due to the fact that the expected spread in orientations
under instrumental noise depends in a complicated way on the
distribution of $|\vec{p}|$. Instead, we used the method of
eigenanalysis described in Sect.\ \ref{sec:pca}.  In our case the
dispersed pulsar signal contributes at most about one fifth of the
total system temperature, justifying the use of a single,
longitude-independent correction to the {\changed spectral density
tensor} for the off-pulse noise, under the caveat that a very small
($\sigma\simeq 0.003$ normalized flux units) amount of measurement
noise contaminates the variances for the central component.  {\changed
The characterstic frequency corresponding to scintillation on the
diffractive time scale (148~s; \citealt{cor86}) is $\sim 1/200$ cycles
per period. To ensure the response was eliminated over its full
frequency extent, we excluded elements of the spectral density
tensor with $|k|/N \leq 1/50$ when computing the covariance matrix.
Using power from $|k|/N \leq 1/50$ in the fluctuation
spectrum of the pulse energy, we measure a modulation index of $0.16$
due to scintillation, in agreement with the measurements and empirical
model of \citet{cor86}, given our observing band. This value
was used to correct the overall scale of the covariance matrix. The
results of the eigenanalysis of this matrix} are shown in
Fig. \ref{fig:pca}.

Beginning with the polarization
entropy (Eq.\ \ref{eq:entropy}), we see that the polarization is most
disordered in the central component, as one might expect from the
distributions seen in the previous sections, but still shows
detectable entropy in all other pulse longitudes accessible to
measurement. That the divergence from pure OPM is significant is
confirmed by the fact that significant variance is detected in the
second and third eigenvalues under every component. The fact that the
second and third eigenvalues are nearly equal in all components except
the central peak indicates that, for these pulse longitudes, the
deviations from OPM show no preferred direction, and cannot be caused
by position angle distortions or a random circular component
alone. This is consistent with the analysis of PSR B1929+10 and PSR
B2020+28 at 1404~MHz performed by \cite{mck04}, who suggests the
superposition of (isotropic) randomly polarized radiation (RPR) as the
cause. On the other hand, in the central component of PSR B0329+54 all
three eigenvalues are significantly different, and indeed the analysis
of the directional distribution in the preceding section shows that
the distribution of $\vec{p}$ cannot be ellipsoidal. This implies that
the deviations are themselves associated with the production of OPM,
as discussed further below.

{\changed It is also interesting to note that there is a suggestion of
correspondence between transitions in the mean modal dominance
(e.g. Fig.\ \ref{fig:anghist}) and local minima in the polarization
entropy (Fig.\ \ref{fig:pca}). This appears to be the case around
pulse longitudes $\sim 42\degr$, $\sim 52\degr$, and $\sim 63\degr$,
however the correspondence is not exact, particularly in the leading
component.  If the trend is in fact real and confirmed in other
pulsars, it would require that any mechanism for the production of two
OPMs predicts that under conditions leading to OPMs of similar
intensity, OPM-related fluctuations dominate more strongly over the
randomly polarized fluctuations than elsewhere in the pulse profile.
This could be the case if the modal intensities tend to be more
variable or more negatively correlated (Eq.\ \ref{eq:sigmasqm}),
and/or  the randomly polarized component is weaker or less
isotropic. A detailed study of a larger sample of pulsars would be
necessary to distinguish between these possibilities.}

We also note that a smooth position angle curve can be constructed
from the eigenvector corresponding to the largest eigenvalue, in
contrast to the position angle of the average polarization vector,
which shows gradual $90\degr$ transitions rather than sharp flips as
would be needed for reconstruction of a continuous smooth curve.  Also
the ellipticity angle curve of the first eigenvector avoids the
problem seen at longitude $\sim 39\degr$, where near complete
cancelling of the linear contributions of the OPMs is not accompanied
by cancelling of the circular component, giving rise to a ``spike'' in
$\chi$ where $\left<\vec{p}\right>/\left|\left<\vec{p}\right>\right|$
sweeps over the left-circular pole (indicating, incidentally,
non-orthogonal modes, or a consistent, superposed left-circularly
polarized component).  These properties will likely make eigenanalysis
a useful technique for detecting polarization fluctuations driven by
OPMs and determining the longitude dependence of their polarization
orientations, even when the signal-to-noise ratio is insufficient to
detect individual pulses.

\subsection{Interpretation of Strong Deviations from OPM}
\label{sec:interp}
The results of the preceding sections show that, while at some
pulse longitudes the circular polarization is typically in proportion
with the linear polarization such that two clusters are produced
in the orientation distribution at antipodal points on the
Poincar\'{e} sphere, at no pulse longitude is the scatter in orientations
consistent with instrumental noise alone. Moreover, near the peak
of the average profile, one mode shows extreme divergence from
the expected orientation, with a correlation between position angle
and ellipticity that is complex in form. The shape is consistent with
a broad, incomplete annulus centred upon the point opposite to
the orientation of the other mode. 

A possible origin for this behaviour lies in birefringent effects in
the magnetosphere. Specifically, the annular form is suggestive of a
propagation effect whereby the polarization state of incoming rays as
represented on the Poincar\'{e} sphere are rotated by a time-varying
angle about the axis defined by the central point of the annulus. Such
an effect is expected if the observed radiation passes through a
region of plasma where the natural propagation modes are different to
the ray polarization, and a net phase delay occurs between the
components of the electric field in each of the two modes due to their
different group velocities. This effect has been termed Generalised
Faraday Rotation (GFR; \citealt{km98}) and is familiar from, but not
theoretically limited to, ordinary Faraday rotation in the
interstellar medium (about Stokes $V$, due to the circular modes of
cold, non-relativistic magnetised plasma), and from the effect of
retardation plates (rotation of $\vec{p}$ about a linear orientation
defined by the optical axis of the material).  In the case of the
pulsar, the polarization of the plasma modes can be identified with
the slightly elliptical polarization states that appear antipodal on
the Poincar\'{e} sphere at the centre of the annulus and at the
typical orientation of states dominated by the other, well-behaved
mode. The incoherent superposition of radiation in the other
polarization mode, which apparently does not suffer this effect, would
cause the annulus to broaden outwards, helping to explain the spread
of the observed distribution.

This kind of effect was predicted for pulsar magnetospheres by
\citet{cr79} and given a quantitative treatment by \citet{lp99}. In
their formulation the change in polarization is effected in the 
{\changed vicinity of}
the polarization limiting region (PLR; Sect.\
\ref{sec:intro}), where the plasma density is
insufficient to cause total decoherence of the modes, yet densities
are still high enough to cause significant phase delays between the
modes. Radiation enters this region as an incoherent mixture of the
local plasma modes, but due to changes in the modal orientation along
the ray path caused by the rotation of the magnetosphere, each of the
incoming rays acquires components in both of the propagation modes,
which propagate at different speeds and alter the polarization of the
ray accordingly. This picture deviates from our observations in
several ways. Firstly, \citet{lp99} assume linearly polarized
propagation modes, whereas the observations indicate elliptical modes,
implying plasma with a net charge density rather than a pure pair
plasma. Secondly, the predicted effect is not as simple as rotation
about a given axis, for the modal polarization orientation, and thus
the axis of rotation, varies along the ray path. {\changed The presence of an
annular shape, as expected from a near-constant polarization of the
propagation modes, may therefore place some constraints on the size of
the region of the magnetosphere contributing significant, variable
amounts of GFR.}
Finally, the effect should only be capable of inducing one
sense of circular polarization \citep{rr90,lp99}, and should affect
both rays equally apart from a reversal in the sense of circular
polarization \citep{pet01}.  

An alternative cause of the misalignment of the polarization of the
propagation modes and the incoming rays, is refraction. \citet{pl00}
show that, while the extraodinary mode propagates under a vacuum
dispersion law, the ordinary mode can suffer from significant
refraction, which, under a non-axisymmetric plasma distribution, can
cause it to move out of the plane of the magnetic field line from
which it originated (and obtained its initial polarization).  The
calculations of \citet{pl00} show that the subsequent alteration of
the polarization state at the PLR can produce either sense of circular
polarization, as seen in our observations. Moreover, since the
extraordinary mode is immune to refraction, it should not suffer the
same PLR effects, consistent with the tight, centrally peaked
distributions of orientations observed here in mode 1. {\changed
Should one of the modes be produced by conversion from the other, as
in \citet{pet01}, this would imply that refraction occurs {\it above}
the conversion region. An alternative means of producing
mode-dependent PLR effects is the invocation of [anti-]correlation
between the efficiency of conversion and the physical conditions in
the PLR \citep{pet01}, however for this to be the case the correlation
must be very strong, given the complete absence of an annulus in the
distribution of states apparently dominated by mode 2.}  That
refraction-driven PLR effects only occur close to the magnetic axis
\citep{pl00} is another prediction borne out by the observation that
only the central profile component shows the annular distribution.
Many pulsars show strong mean circular polarization in central,
so-called ``core'' components \citep{ran83}, which tends to show a
central sense reversal (\citealt{rr90}, although see also
\citealt{hmxq98}) taken by \citet{pl00} as support for their model of
the refraction-driven PLR effect. The probable direct detection of
this effect in PSR B0329+54 opens the possibility of good tests of the
model through applications of the techniques used here on a larger
sample of pulsars with and without ``core'' components, and
examination of the frequency dependence.

\begin{figure}
\resizebox{0.9\hsize}{!}{\includegraphics{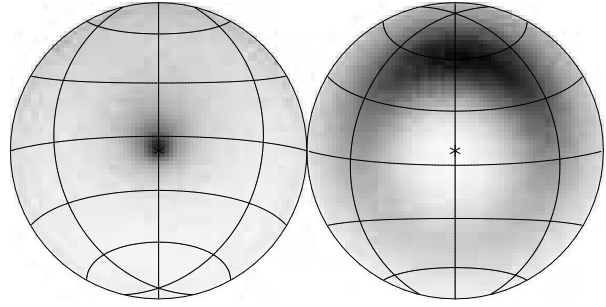}}
\caption{Distribution of polarization orientations deriving from
a simulation involving GFR (see text). The projection used is as
in Fig.\ \ref{fig:po}, plotted in a linear grey density scale.}
\label{fig:sim}
\end{figure}

Detailed modelling of this effect is beyond the scope of this work,
however to prove the basic assertion that GFR can produce the spread
of orientations observed, we have performed some basic numerical
simulations.  We simulated the observed polarization vector as the
sum of three components, the ordinary ray,
the extraordinary ray  and an RPR component:
\begin{equation}
\vec{p} = \vec{p_{\rm o}} + \vec{p_{\rm e}} + \vec{p_{\rm r}}.
\end{equation}
\noindent {\changed We assume that certain parameters are independently
distributed according to chosen ``reasonable'' distributions. These
are merely intended to provide variation in order to produce the
qualitative features of the model, and are not expected to necessarily
closely resemble the distributions under a complete physical model.}
The ordinary mode was presumed to have an initially linear
polarization orientation with a position angle drawn from a von Mises
distribution:
\begin{equation}
f(2\psi) = \frac{e^{\kappa\cos (2\psi-\mu_{\rm v})}}{2\pi I_0(\kappa)}
\end{equation}
(where $I_0$ is a modified Bessel function of the first kind)
with $\mu_{\rm v}=50\degr$, $\kappa=20$, while its 
(polarized) intensity was drawn from an exponential distribution:
\begin{equation}
f(I) = \left\{ 
\begin{array}{ll}
\frac{1}{\mu}e^{-I/\mu_{\rm e}}  &   I \geq 0 \\
 0                      &   I < 0 
\end{array}
\right. ,
\end{equation}
\noindent with $\mu_{\rm e}=1$.  {\changed To model the effect of GFR,
the} initial state for the ordinary mode was rotated by an angle
$\theta$ about an axis $\vec{1_{\rm m}}$ oriented with $\psi=90\degr$,
$\chi=4\degr$, where $\theta$ was also drawn from a von Mises
distribution ($\mu_{\rm v}=90\degr$, $\kappa=1$), to give the computed
$\vec{p_{\rm o}}$.  The extraordinary mode ($\vec{p_{\rm e}}$) was
presumed to be always oriented with $\psi=0\degr$, $\chi=-4\degr$,
with a (polarized) intensity drawn from an exponential distribution of
mean $\mu_{\rm e}=0.2$. The contribution of RPR ($\vec{p_{\rm r}}$)
was calculated by drawing its components $Q$, $U$, $V$ from
independent Gaussian distributions:
\begin{equation}
f(x) = \frac{1}{\sigma\sqrt{2\pi}}e^{\frac{(x-\mu_{\rm g})^2}{2\sigma^2}},
\end{equation}
with $\mu_{\rm g}=0$ and $\sigma=0.05$.  This model was realised $10^6$ times
to produce the distribution of orientations depicted in Fig.\
\ref{fig:sim}. Although clearly the chosen distributions are not
perfect, the resemblance to the observed distributions is
striking, confirming the possibility that GFR is in effect.

\section{Conclusions}
We have performed a detailed analysis of the polarization of
individual pulses from PSR B0329+54 at 328~MHz. The use of statistics
involving the ellipticity angle instead of Stokes $V$ or $V/I$ enabled
us to reveal for the first time the rich structure of the circular
polarization distribution, which is correlated with the position angle
distribution. 

Displaying the distribution of polarization orientations
in an equal-area projection, a remarkable structure was uncovered for
pulse longitudes near the centre of the average pulse. The
annulus-like form of the distribution in one mode is, in our view,
indicative of Generalised Faraday Rotation (GFR) in the pulsar
magnetosphere, while the fact that it apparently only affects one
polarization mode, and is capable of producing either sense of
circular polarization, is taken as an indication that refraction of
the ordinary mode is taking place. The apparent ellipticity of the
polarization state upon which the annulus centres and orthogonal to
the typical polarization of states dominated by the other mode are
taken to indicate that the GFR occurs in a region with elliptically
polarized plasma propagation modes, indicating a net charge density
in the plasma. 

Through the analysis of the covariance of the Stokes parameters,
by means of eigenvector decomposition, we have shown  quantitatively 
for the first time that  the circular polarization is consistent with
an origin in elliptically polarized orthogonal polarization modes.
Moreover, we find that a significant apparently randomly polarized
component dilutes the purity of OPM-driven variations in polarization
state, as also found recently by \citet{mck04} for  PSR B1929+10 and
PSR B2020+28 at 1404~MHz. 

In this work we have shown how to detect and characterise deviations
from OPM in more powerful ways than previously available. Future
application of these techniques to other pulsars over a broad
frequency range should allow renewed progress in the
resolution of decades-long debate about the origin of pulsar
polarization.

\acknowledgements The authors thank W. van Straten for fruitful
discussions. {\changed For their useful comments on the manuscript we
thank S. Petrova, and the referee who raised issues that led to
important improvements. } RTE is supported by a NOVA fellowship.
The Westerbork Synthesis Radio Telescope is administered by ASTRON
with support from the Netherlands Organisation for Scientific
Research (NWO).

\bibliographystyle{aa}

\appendix
\section{Polarization Calibration}
The Westerbork Synthesis Radio Telescope (WSRT) is an array of 14 dishes,
each of two nominally linear dipoles, which is configured as a tied
array during pulsar or VLBI observations.  Assuming that the response
of each telescope to the incident electric field is linear, the
observed voltages can be described using a Jones matrix \citep{jon41}
for each dish, $i$:
\begin{equation}
\vec{v_i} = \vec{J_i}\vec{e}.
\end{equation}
\noindent The procedure for forming a tied array is to add the voltages
of all telescopes after correcting them for relative phase (due
to the source direction):
\begin{equation}
\vec{v_{\rm tot}} = \sum_{i=1}^{N} e^{-i\theta_i}\vec{J_i}\vec{e}.
\end{equation}
\noindent where $N$ is the number of telescopes. It is therefore
trivially true that the tied array signal can be calibrated by
determining an overall system Jones matrix, i.e.:
\begin{eqnarray}
\vec{v_{\rm tot}} &=& \vec{J_{\rm tot}}\vec{e}, \\
\vec{J_{\rm tot}} &=& \sum_{i=1}^{N} e^{-i\theta_i}\vec{J_i}.
\label{eq:jonessum}
\end{eqnarray}
The problem of tied array calibration is thus no different to the
calibration of single-dish observations, for which several recent
treatments are available that avoid potentially dangerous simplifying
assumptions of earlier works
\citep{hbs96,shb96,ham00,bri00,van02,joh02,van04}.  These approaches
are based on the Jones matrix formalism, which has seven degrees of
freedom in the instrumental model of each frequency channel (the
eighth, absolute phase, is lost in the computation of the self
coherency matrix for single dish or tied array
observations). \citet{ham00} divides these into a ``polrotation'',
which causes a pure rotation of $\vec{p}$ about an arbitrary axis by
an arbitrary angle (3 degrees of freedom), a ``poldistortion'', which
exchanges an arbitrary amount of power between Stokes I and and an
arbitrarily oriented component of $\vec{p}$ (3 degrees of freedom),
and an overall system gain (one degree of freedom).  A similar
decomposition was made independently by \citet{bri00}, who notes that
Jones matrix transformations are isomorphic with the Lorentz group,
with the type of transformation called a ``poldistortion'' by
\citet{ham00} corresponding to a Lorentz boost.  Apart from providing
a convenient geometric interpretation of the kinds of transformations
effected by linear system components, as \citet{ham00} notes this
decomposition suggests a calibration strategy that begins with the
observation of a presumed unpolarized source, immediately giving the
gain and poldistortion. The polrotation can then be obtained by
observations of a further two sources about which some assumptions can
safely be made (for example, that they are linearly polarized and/or
have a given position angle).

Unfortunately efforts to calibrate the WSRT in tied array mode are
severely hampered by the inclusion of non-linear system components.
Specifically, the voltage signals from each telescope are sampled
using thresholds that are dynamically determined using a control
system that attempts to maintain the variances of the sampled signals
at a constant level, based on averaging with some time constant of the
order 1 ms. These are known as the automatic gain controllers
(AGCs). Although these components might at first appear disastrous for
the detection of non-stationary signals, the fact that they normalise
the individual telescope powers while the source signal adds
coherently means that the distortion is reduced for $N\gg
1$\footnote{See the PuMa manual, \\
{\tt http://www.astron.nl/wsrt/WSRTobs/PuManual.pdf}}. However, the fact
remains that the AGCs introduce an unknown, time-varying gain to each
of the real and imaginary parts of each polarization channel of each
telescope, making precise calibration impossible for tied array
observations, and we strongly suggest that future telescope arrays
provide tied-array systems free of AGCs. Since the system is not
precisely calibratable, we limit the scope of our calibration to the
grossest of instrumental effects, differential gain and phase between
the summed X and Y polarization channels.  Remaining effects are
expected in any case to be small, due to the high accuracy of dipole
setting at WSRT \citep{ww77}, and the tendency for random error terms
to cancel in the sum of Eq.\ \ref{eq:jonessum}.  By comparison of our
observations with published polarimetric profiles available via the
EPN
database\footnote{\tt http://www.mpifr-bonn.mpg.de/div/pulsar/data/},
we estimate that our results are accurate to within a few percent,
certainly sufficient for examination of the basic polarization
properties of the source.

The procedure we use for determining the differential phase and gain
is based on the differential Faraday rotation across the observing
band. This provides a variety of input polarization states incident on
the telescope, with a known relationship between them, allowing one to
fit simultaneously for the source polarization and the instrumental
response. In this regard the technique is similar to previous
strategies employing the parallactic angle rotation during long
observations of a source \citep{scr+84}, which cannot be used at the
WSRT because its equatorial mounts cause the dipoles to track the
parallactic rotation (by the same token, this eliminates a potential
source of time-variability in incompletely calibrated
measurements). Just as the parallactic technique is limited to the
assumption that neither the source polarization nor the telescope
response changes with time, our method assumes frequency independence
except for an overall factor incorporating the gain and intrinsic
intensity spectrum of the source. The technique is also limited by the
commutativity of certain transformations with the known transformation
effected by Faraday or parallactic rotation, about Stokes $V$.  That is, 
for any
solution of the system Mueller matrix (which is constructed from
the Jones matrix, see e.g. \citealt{hbs96}) satisfying
\begin{equation}
\vec{S_i}' = \vec{M} \vec{R_V}(2\theta_i)\vec{S},
\end{equation}
where $\vec{S}$ is the intrinsic Stokes 4-vector, $\vec{S_i}'$ is a
measured (uncalibrated) Stokes 4-vector observed under a known Faraday
or parallactic rotation angle $\theta_i$ and $\vec{R_V}(2\theta_i)$ is
a Mueller matrix effecting a rotation of angle $2\theta_i$ about
Stokes $V$, there exists an infinite number of other solutions
\begin{equation}
\vec{S_i}' = \vec{M}\vec{E}\vec{R_V}(2\theta_i)\vec{E}^{-1}\vec{S},
\end{equation}
where $\vec{E}$ is any matrix that commutes with
$\vec{R_V}(2\theta_i)$. Without additional constraints, this approach
is therefore only able to determine the system response to within an
arbitrary rotation (``polrotation'') and boost (``poldistortion'') in
Stokes $V$. However, as stated above we have chosen to use a reduced
description of the system response, parameterised only by
frequency-dependent gain, and X-Y differential gain and phase.  For
assumed perfect linear dipoles, differential phase and gain errors
correspond to rotations and boosts about Stokes $Q$
(e.g. \citealt{bri00}), which do not commute with $\vec{R_V}$,
obviating the need for additional constraints.  Our model is thus:
\begin{equation}
\vec{S}'(\nu) = g(\nu) 
\vec{R_Q}(2{\rm RM}(c/\nu)^2+\theta)
 \vec{B_Q}(\beta) \vec{S} + {\rm noise},
\end{equation}
\noindent where $g(\nu)$ models the frequency dependent gain and
source spectrum, {\rm RM} is the rotation measure of the propagation
path between the source and Earth, $\theta$ is the phase offset between
X and Y channels, and $\beta=\ln g_x/g_y$ where $g_x$ and $g_y$ are
the (voltage) gains of the X and Y channels. 

\begin{figure}
\resizebox{0.9\hsize}{!}{\includegraphics{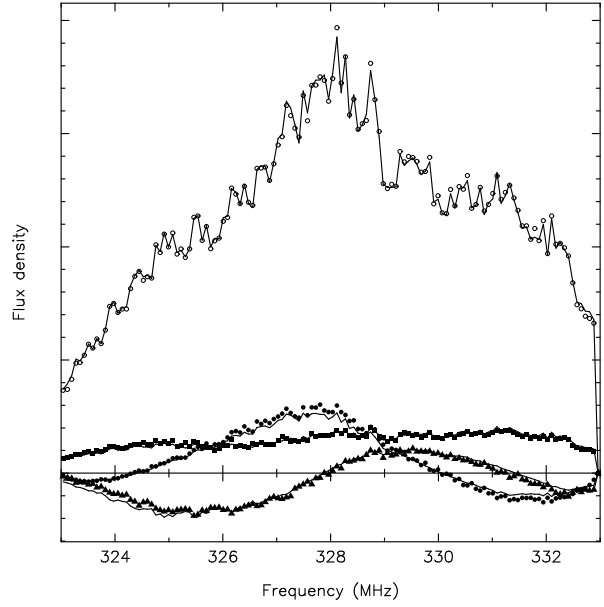}}
\caption{Observed values of Stokes $I$ (open circles), $Q$ (triangles),
$U$ (squares) and $V$ (filled circles) in the peak longitude bin of
PSR B0329+54, as a function of frequency. Predictions of the best-fitting
model (incorporating all longitude bins) are shown with lines.}
\label{fig:fcal}
\end{figure}

We define the best-fit system response as the one that minimises the
global $\chi^2$:
\begin{equation}
\chi^2 = \sum_i \left(
\frac{\vec{D_i}_{I}^2}{\sigma_{I,i}^2} +
\frac{\vec{D_i}_{Q}^2}{\sigma_{Q,i}^2} +
\frac{\vec{D_i}_{U}^2}{\sigma_{U,i}^2} +
\frac{\vec{D_i}_{V}^2}{\sigma_{V,i}^2} 
\right),
\label{eq:chisq}
\end{equation}
\noindent where $i$ indexes all measured Stokes vectors (i.e.  all
frequency channel/pulse longitude pairs), $\sigma^2_{\{IQUV\}}$ are the
variances of each of the Stokes parameters in off-pulse regions
(potentially also dependent on $i$), and
\begin{equation}
\vec{D_i} = \vec{S}'_i - \vec{M_i}\vec{R_V}(2\theta_i)\vec{S},
\end{equation}
\noindent where in our case, $\vec{M_i}=g(\nu_i) \vec{R_Q}(2{\rm
RM}(c/\nu_i)^2+\theta) \vec{B_Q}(\beta)$. In the general case where a
full system calibration is achievable $\vec{M}$ should be parameterised as a
function of frequency by the seven real and imaginary components of
the elements of the Jones matrix (the imaginary part of one of the
diagonals can be set to zero since absolute phase is irrelevant).
\citet{joh02} describes a similar scheme to that described above,
employing a Jones matrix formalism and parallactic angle rotation,
from which we differ in several respects. Firstly rather than
minimising $\chi^2$, which corresponds to the maximum likelihood
estimator, \citet{joh02} minimises the Manhattan distance, which is
likely to have poorer estimation accuracy than maximum
likelihood. Secondly, \citet{joh02} discusses the difficulties of
multi-step calibration schemes but does not propose an
alternative. Including all measurements of all sources in the
global $\chi^2$, including any injected noise sources and continuum
calibrators, and minimising it in one step is in our view the best way
to avoid these difficulties. Finally, we caution against neglecting
second-order terms in the expansion of the Mueller matrix as it is
unnecessary and introduces the possibility for error if and when the
technique is applied to telescopes with non-negligible cross terms
(e.g. LOFAR/SKA). {\changed We note that the issues described
above have received independent treatment in the recent work of
\citet{van04}, who also provides a method for minimization of
Eq.\ \ref{eq:chisq} that avoids explicit expansion of the
Mueller matrix and its partial derivatives.}

We performed the procedure described above using a frequency-resolved
polarimetric profile from the first 12500 pulses of the observation of
PSR B0329+54 described in the text. That $\theta\simeq \pm90\degr$ is
clearly visible in the frequency dependence of the Stokes parameters
(Fig.\ \ref{fig:fcal}), where the Faraday modulation appears mainly in
$Q$ and $V$, instead of $Q$ and $U$ as expected if $\theta=0$. In fact
it was this feature, seen in this and other WSRT observations in the
328~MHz band that led us to the calibration scheme described here. The
results of the fit support the assertion that $\theta\neq0$, yielding
$\theta=-75.24\degr$ and $\beta=-0.0278$, implying $g_x/g_y=0.973$
(the model predictions using these values and the other parameters are
plotted in Fig.\ \ref{fig:fcal}). These values were used to correct the
recorded Stokes vectors before further analysis described in the body
of this paper.

\end{document}